# Flexible and High-Performance Radio Access Networks for upcoming Sixth-Generation (6G) Systems

Peter Schefczik[1], Umar Toseef[1], Paolo Baracca[2], Ralf Klotsche[1], Torsten Dudda[3], Mai-Anh Phan[3], Lorenzo Miretti[4], David Ginthoer[5], Bin Han[6]

[1]Nokia Solutions and Networks GmbH & Co. KG, 70469 Stuttgart, Germany
[2]Nokia Solutions and Networks GmbH & Co. KG, 81541 München, Germany
[3]Ericsson GmbH, 52134 Herzogenrath, Germany
[4]Fraunhofer Heinrich Hertz Institute, 10587 Berlin, Germany
[5]Robert Bosch GmbH, 70469 Stuttgart, Germany
[6]RPTU Kaiserslautern-Landau, 67663 Kaiserslautern, Germany

"This research work was supported by the German Federal Ministry Research, Technology and Space (BMFTR) as part of the project 6G-ANNA. However, the authors alone are responsible for the content of this paper."

**ABSTRACT** The collaborative research project 6G-ANNA develops concepts for the 6G radio access network (RAN) architecture and technology components. Previous RAN generations have become inherently more complex and reach their limits in handling foreseen future traffic demands with their diverse characteristics in an efficient manner, e.g., for the use-case of mobile eXtended Reality (XR) on a massive scale. One main objective of 6G is to regain both operational and energy efficiency, i.e., by simplification and automation. To achieve this, in this paper a flexible 6G RAN functional architecture and protocol stack as well as implementation and deployment options are described. We outline how performance is optimized by distributed Multiple Input Multiple Output (MIMO) and distributed Carrier Aggregation (CA), and furthermore, how adaptiveness and scalability is enabled by Cloud RAN and service orchestration. Finally, the proposed zero-trust framework mitigates security risks in the described 6G RAN architecture.

## I. INTRODUCTION

The unprecedented data growth, the ubiquity of the mobile data network and the emergence of new, demanding use cases necessitate a paradigm shift in the mobile network infrastructure. This means that challenges in performance, flexibility, efficiency and security must be addressed for 6G to succeed. Metaverse applications of shared virtual and digital worlds with massive user capacity, real-time interactions, and seamless integration of physical and digital world is one of such upcoming use cases. These include also mobile eXtended Reality (XR), which is considered a driving use-case for 6G [1]. Mobile XR refers to providing high-quality digital experience to end-consumers, involving multiple human senses. It is an evolution of the enhanced Mobile BroadBand (eMBB) data and voice communication and further develops the XR use case considered in 5G-Advanced. Wide area coverage and capacity are offered to support end-consumers wearing Head-Mounted Displays (HMDs) and potentially sensors and actuators. The end-consumers can move freely outdoors, indoors, and in vehicles, while the network enables seamless connectivity. The requirements for mobile XR in the 6G timeframe are largely dependent on the foreseen applications, like healthcare, robot controls or just remote presence. It is meaningful that the 6G RAN aims to support at least the XR requirements agreed for 5G-Advanced RAN, while also supporting more users and achieving higher energy efficiency compared to 5G-Advanced.

The potential benefits of cloud-based RAN, such as improved scalability, flexibility, and cost-efficiency, motivate the cloud-based RAN architecture research on the 6G RAN architecture design, which is one of the working tasks in the German lighthouse project 6G-ANNA [2]. The cloud-based architecture is a fundamental component of the design of the 6G RAN by 6G-ANNA. Moreover, the concept of network slicing for the User and Control plane components in the RAN is introduced in the 6G-ANNA project. Within the cloud-based RAN, network slicing and flexible function placement work together to achieve higher pooling gains and energy efficiency while meeting the service requirements. In [3], the Architecture Working Group within the Smart Networks and Services Joint Undertaking (SNS JU) initiative provides a comprehensive overview of key technology enablers and cutting-edge design trends shaping the 5G architecture. The work also explores promising trends towards the evolution of 6G.

In [4], the authors envision a heterogeneous and distributed cloud environment for the upcoming 6G communication system, where resources are spread across multiple clouds in various locations. The motivation for advancing network slicing to a fully slice aware network for 6G, followed by an exploration of enabling technologies such as virtualization, softwarization, and cloudification are discussed in [5].

In the recent 3GPP 6G WS, as summarized in [6], first visions on 6G architecture by operators and vendors were presented. A trend towards simplification of architecture and protocols and their configurability options, compared to 5G, as well as a preference for 6G standalone operation were noted. However, with the discussions just being started, it is still unclear how a to be specified 6G architecture would eventually look like.

This paper explores innovative RAN architecture aspects, drawing upon the current understanding of 3GPP mobile network generations up to 5G. Such aspects are e.g., the dynamic network function placement and migration. These challenges have also been considered within the broader context of the 6G End-to-End (E2E) architecture along with a functional view of the 6G reference architecture presented in [7]. Moreover, our proposed flexible 6G RAN architecture overcomes existing problems of the current 5G architecture, such as the multitude of architectural choices, including spectrum aggregation options, higher layer split, and distributed packet processing. These architectural choices have resulted in increased complexity, higher processing demands, and costs, and thus have delayed the market rollout of 5G. Our main objective is to overcome these problems in the upcoming 6G RAN architecture.

The rest of this paper is organized as follows. In Sect. 2, we delve into the limitations of the current 5G RAN architecture, highlighting key issues such as too many architecture options, too distributed UE knowledge, too many retransmission options all impacting the flexibility of the 5G RAN. Sect. 3 elaborates on the key design goals for the 6G RAN, including AI and cloud-native capabilities, energy efficiency, and trustworthiness. We also examine implementation and deployment flexibility, focusing on joint resource scheduling, and dynamic function placement. In Sect. 4, we depict potential solutions for achieving these 6G RAN design goals, encompassing a range of methodologies, including network slicing for the RAN, protocol stack optimizations, carrier aggregation possibilities in 6G, a novel sub-network concept and a zero-trust access control. Finally, Sect. 5 concludes by summarizing the key findings and outlining potential directions for future research, including the application of the service-based architecture concept also inside the RAN.

## II. ISSUES IN CURRENT 5G RAN

In this section we discuss drawbacks of the current 5G RAN architecture specification and deployments that limit flexibility and need improvements in e.g., energy-efficient processing, signaling and RAN protocol stack complications.

### A. MULTIPLE ARCHITECTURE OPTIONS

5G NR specified different architecture options for spectrum aggregation, such as carrier aggregation (CA) and dual connectivity (DC). In CA radio resources are aggregated by a single gNB and downlink CA doesn't imply also uplink CA, i.e., UL power is not necessarily split among carriers. In DC, on the other hand, the resources are aggregated by different gNBs with their respective schedulers via an inter-gNB interface. Also, an UL connection is required to both gNBs. For DC, different protocol stack sub-options exist on how to split data among the transmission paths utilizing different frequency carriers. These various options created complex specifications, fragmented introduction of NR, and delayed introduction of 5G NR standalone to the market.

### B. F1/E1 INTERFACES INCREASE SIGNALING DELAY

5G NR introduced the higher layer split, see Fig. 1, which divides the gNB into two parts (with a multi-vendor interface between them), i.e., a centralized unit (CU) and a distributed unit (DU), respectively. This substantially complicated the decision logic and overall complexity in both the CU and the DU while leading to worse decisions (e.g. which UE capabilities and resources to utilize, based on distributed information), delayed actions (due to latency over the interfaces) and increased processing load compared to a monolithic gNB architecture without any split [8]. Similarly, for the user plane, packet processing and buffering in 5G NR is split between CU and DU, requiring duplicated memory accesses and CPU cycles without performance benefits. Distributing pending data between CU and DU further required a flow control protocol between the nodes. This way, active queue management (for interaction with transport layer protocols to keep the transmit buffer latency

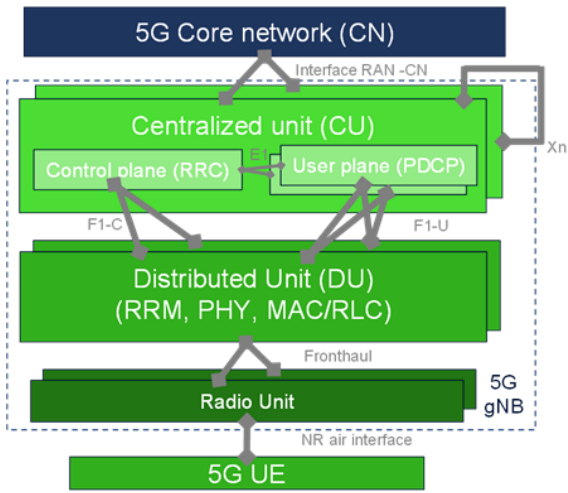

FIGURE 1. 5G RAN with CU/DU split.

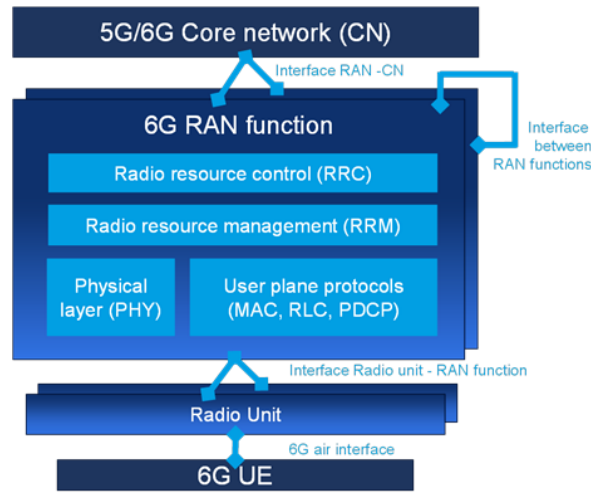

FIGURE 2. 6G RAN architecture proposal.

to a minimum, while still maximizing the link capacity) becomes inherently complex.

### C. RETRANSMISSION PROTOCOL OPTIONS

The protocol stack in 5G NR is organized in a way that supports DC, e.g., with reordering among the transmission paths, where long timer settings may lead to latency spikes. When it comes to retransmission protocols in 5G NR, we observe that these are currently spread over three different layers:

- Retransmissions using hybrid automated repeat request (HARQ), based on physical layer feedback, to combat residual errors from forward error correction over the air as fast as possible (MAC).
- Retransmissions to correct residual errors from HARQ, in a reliable way, i.e., as automated repeat request (ARQ) on radio link control (RLC) layer.
- Retransmissions to recover from losses during handover interruption and related lower layer protocol resets, on packet data convergence protocol layer (PDCP).

### D. LIMITED FLEXIBILITY

Traditionally, RAN deployments have relied on a fixed architecture tailored to anticipated traffic patterns within a predetermined radio deployment. Lower-layer processing functions are still implemented on specialized hardware, limiting flexibility for feature additions or future upgrades. Also, this approach struggles to adapt to evolving business needs, emerging technologies, and the growing trend towards cloud-native implementations on Common Off The Shelf (COTS) hardware within data centers. For example, the rise of Mobile XR as a driving use-case for 6G presents significant challenges for the traditional RAN network infrastructure. All those limitations led us to specific design goals, which are outlined in the following section. These goals encompass not only simplification, AI and cloud-native capabilities, but also energy efficiency, trustworthiness, deployment flexibility, network slicing, joint resource scheduling, and dynamic function placement.

### III. 6G RAN ARCHITECTURE DESIGN GOALS

In this section, we explore the design goals of the 6G RAN architecture. Beyond simplification, AI and cloud-native capabilities, energy efficiency, and trustworthiness, we also delve into implementation and deployment flexibility. This includes network slicing implication on RAN, joint resource scheduling, and dynamic function placement.

#### A. FLEXIBILITY BY DECOMPOSITION AND SLICING OF UE CONTROL AND USER PLANE FUNCTIONS

Building on the 5G RAN decomposition in 6G, i.e. namely separating control and user plane into service specific components, enables independent scaling of resources, enhancing security and redundancy through physical isolation of network slices or services. This approach optimizes for specific service types, meeting low latency requirements, supporting diverse QoS needs and also allows for "pay as you grow" scaling of the On-Premises (OnPrem) resources. By offloading less delay-sensitive traffic to the more central Far-Edge (FarEdge) Cloud, OnPrem processing is reduced while ensuring low-latency traffic remains local for optimal performance.

#### B. EFFICIENCY BY SIMPLIFICATION

One key design goal for the 6G radio access is to achieve simplicity in its specification and architecture options compared to 5G (see also [6]), i.e., to reduce costs of implementing and maintaining various options. This can be achieved by concentrating on key open interfaces in RAN where their complexity is proven to bring a functional or performance benefit. Multiple redundant features or solutions (e.g., for spectrum aggregation) should be avoided, because these increase complexity, hinder interoperability and lead to overly complicated standards.

### C. DYNAMIC FUNCTION PLACEMENT

Dynamic function placement in the RAN allows for optimal resource utilization by placing RAN processing functions closer to the RU when needed by the service, thereby improving performance and reducing latency. Such flexibility is enabled by virtualization and decoupling software functions from specific hardware. Furthermore, dynamic function placement can adapt to changing network conditions and traffic patterns, thereby ensuring optimal network performance through load balancing.

### D. SLICING AND JOINT RESOURCE SCHEDULING

The new architecture of the 6G RAN with its novel features for dynamic function placement, scaling, and network slicing raises new opportunities for the scheduling of air interface resources. Each slice and even each service of an individual UE have requirements on how to handle the traffic, but finally they all compete for a limited number of processing resources for user plane and control plane processing. The new services of 6G require a fast reaction (in real time) on transmission requests and the network slices require flexibility in placement and service handling, and the desired spectral efficiency requires scheduling over the whole set of resources. Thus at least one stage of scheduling needs to be at a central unit, which balances all requests at a given time over the available resources. The scheduler for a RAN is not standardized and thus it is an opportunity for each supplier to implement its own strategy. The challenge for a good scheduler design is to distribute and delegate pre-processing steps into the network functions of the slices and minimize the number of final processing steps, which are needed to be executed in one place, while providing the service level agreements (SLA), a good spectral efficiency and energy savings.

### E. ENERGY EFFICIENCY

The above design goals also significantly contribute to achieving the energy efficiency targets of the 6G RAN. In particular, this facilitates the introduction of sleep modes with different time granularities for various network components according to specific needs [9]. Additionally, from a 6G RAN perspective, energy efficiency can be further enhanced by supporting more advanced radio access technologies, such as those based on distributed MIMO techniques, described later in Sect. IV-G.

### F. TRUSTWORTHINESS

Trustworthiness is a critical objective in 6G networks, encompassing security, privacy, and resilience to ensure reliable and confident mobile communication services. In the complex 6G ecosystem, managing trust is challenging. Trust relationships cannot be propagated, inherited, or transferred among stakeholders, and they are not necessarily reciprocal. Misunderstandings and mismanagement of these relationships, however, are common in practice, which increase risks and vulnerabilities of the 6G system. To mitigate these risks, the 6G network should adopt a zero-trust framework, ensuring users can trust the network while the network does not inherently trust other parties. Achieving trustworthiness within this framework requires robust mutual AAA mechanisms among various stakeholders. Continuous assessment and management of trustworthiness, along with comprehensive auditing mechanisms, are essential to maintaining security and resilience. Integrating these principles into the 6G RAN architecture establishes a robust trustworthiness framework, ensuring secure, private, and resilient communication, while meeting the high expectations of future mobile communication systems.

## IV. SOLUTIONS FOR 6G RAN

In order to fulfill the design goals for 6G RAN as outlined in the section above, we describe key solution approaches in the following. In the design for the 6G system architecture, it is important to distinguish between different aspects that define the architecture. The complete 6G system will be made up of a combination of functional, implementation, and deployment architectures.

### A. FUNCTIONAL ARCHITECTURE ENHANCEMENTS

The standardized functional architecture abstracts entities and functions, defines interfaces for services offered for other entities, and as well defines the procedures of the system. Thus, it should dictate as little of implementation and deployment as possible. Out of the three architectures, the functional architecture is the one requiring the longest time cycle to change. This longer timeframe, however, allows for greater flexibility and potential innovation in the other two architectures. For example, the functional architecture should be location agnostic as the specific location of the functions should be determined by the deployment architecture. In contrast, the 5G standard provides details and explicit support for many implementation and deployment options. The functional architecture standards should ideally break away from the notion of a node and should focus on logical network functions. Handling of these network functions should be de-coupled from the hardware and software implementing them. This abstracts away any deployment choices, or infrastructure/Cloud implementation details, i.e., allows freedom in implementation and de-couples, e.g., standards development cycles from Cloud-infrastructure development cycles. The implementation architecture provides the structures to build the system as part of the development process. The deployment architecture gives network operators concrete plans on network design and its configuration among the available implementations.
The envisioned functional (to be standardized) 6G RAN architecture with its defined open interfaces is illustrated in

Fig. 2. It is independent of the underlying infrastructure layer, and this way allows flexible implementations, e.g., on a Cloud environment.

The idea of the 6G RAN function (RANF) is that it optimizes the UE's performance in terms of achievable rates and latencies while having low mobility interruptions within a certain geographical area, that is covered by one or multiple connected radio units (RUs), i.e., transmission/reception points [10]. For this purpose, the sub-functions of the 6G RANF must tightly interwork for efficient and high-performance operation based on the shared UE's context, i.e., its capabilities and current state. These sub-functions are, respectively, radio resource control (RRC), radio resource management (RRM), the physical layer (PHY) and the user plane protocols. The 6G RANF may be executed as a function in the Cloud RAN environment. Due to the interdependencies of its sub-functions (shared UE context), an internal split (standardized split for inter-vendor deployment) would lead to additional complexities without an apparent performance advantage. On the contrary, the splitting would lead to additional signaling overhead and latency. Another issue would be the split in the user plane, requiring multiple processing entities and potential buffering points with an additional flow control protocol in-between, i.e., to distribute the adequate amounts of data among these entities. These complexities became also obvious for the standardized F1 interface in 5G, i.e., the so-called higher layer split (HLS), hampering the market success in inter-vendor deployments. This interface is thus not foreseen in the envisioned 6G RAN architecture [8].

In this new architecture, the UE connects to the 6G RANF via the UE-RANF interface, i.e., the radio interface. The UE may utilize one or multiple transmission/reception points provided by the radio units (RUs), which are connected to the 6G RANF via a defined open RU-RANF interface, also called the lower layer split (LLS). This allows an operator to choose optimal RUs, e.g., hardware-optimized for certain frequency ranges, traffic types, and load, independent of the choice for the 6G RANF provider. It is noteworthy that the LLS doesn't come with the complexities associated with the HLS, as the RU is employed mostly as a processing node for the UE, and doesn't require a split of the UE context, capabilities or connection state.

The 6G RANF connects to the evolved 5G/6G core network (CN), via a RAN – CN interface, proven to be a well-established open interface in the market as it clearly separates domains of concern, such as managing and operating resources in RAN, and managing access and services in CN.

Another interface to consider is the RANF-RANF interface interconnecting the RANF, e.g., for mobility purposes. The approach is to keep this interface rather simple and avoid complex functionalities such as DC via this interface as in 5G. More potential is seen in optimizing radio resources, e.g., via CA within a geographical area via potentially multiple RUs connected to the same 6G RANF, rather than on higher protocol layer via different nodes/functions (see also Sect. IV-H).

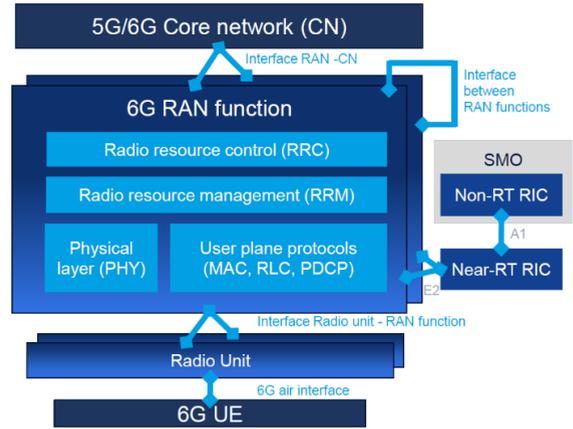

**FIGURE 3.** 6G RAN with RIC.

Implementation and deployment flexibility is achieved in this functional architecture approach based on the focus on these key open interfaces as well as the abstraction from the infrastructure layer below.

In addition to the described functional split, further open interfaces may be defined for automation and management purposes as envisaged in the following section.

### B. O-RAN INTEGRATION

The open RAN concept is based on the disaggregation of RAN components enabled by software virtualization and open interface specifications. It offers greater optimization possibilities by allowing flexible network and service selection, usage of context-information from outside the RAN, and diverse deployment options such as cloud-based.

The open RAN architecture is defined in the O-RAN specification, e.g. in [11]. A key component is the RAN Intelligent Controller (RIC), which implements intelligent and automated control functions that can be AI/ML-based in form of xApps and rApps. 6G will continue to see a plethora of use cases ranging from low latency to high data rate requirements including new concepts such as highly specialized sub-networks. The O-RAN architecture offers programmability of the network via new interfaces such as the E2 over which reports from the RAN components can be gathered and policies can be inserted. This is a big step towards intent-driven and autonomous network configuration, which becomes even more important in 6G to adapt the RAN to the diversified service requirements coming, e.g., from verticals in a timely and cost-effective manner. Applied to our 6G RAN architecture proposal, as in Fig. 3, the RIC would interface with the RANF and Radio Unit, for example.

### C. A SLICE-BASED APPROACH FOR THE RAN ARCHITECTURE

The European flagship project HEXA-X-II [1] has identified six use case families showing the need for a new 6th generation. From the several key performance indicator (KPI) values treated, we focus on latency and reliability to derive RAN QoS classes. These 6G RAN QoS classes range from mission critical with lowest latency (0.1 ms - 10 ms) and highest reliability ($1-10^{-8}$ - $1-10^{-10}$) until moderate latency (20 ms – 1100 ms) and lower reliability ($1-10^{-4}$ - $1-10^{-7}$). These QoS classes are in line with the ITU-R recommendations [12] where over the radio interface latencies are targeted at 0.1 ms – 1 ms and for reliability range from $1-10^{-5}$ to $1-10^{-7}$. To simplify our architecture considerations these RAN QoS classes can be mapped to two different RAN slices. More than two slices are of course also possible.

- RAN slice I, containing all RAN QoS class combinations including the mission critical RAN QoS class, for example URLLC.
- RAN slice II, containing all RAN QoS class combinations for example eMBB, excluding mission critical or very low latency (0.1 ms - 10 ms) classes.

In detailing the high-level architecture of 6G shown in Fig. 2, an example deployment for the two RAN slices is depicted in Fig. 4.

The functional decomposition into slice based 6G RAN modules and their placement possibilities and functionalities are as follows:

- RRM, can be placed OnPrem only and contains the functionalities for Cell-Control Plane (CP), Cell Setup, UE Access procedure per cell, Radio admission control and L2 packet scheduling (L2-PS) per cell.
- RRC, is present per RAN Slice, and can be instantiated on OnPrem and FarEdge cloud. It contains the functionalities for UE-Control Plane with Slice admission control and bearer setup.
- CP-Routing is working on FarEdge and routes Core Network Control Plane messages to RRCs.
- User Plane (UP) is instantiated per RAN Slice and can be placed either OnPrem or on FarEdge cloud. UP is responsible for L2 non real-time and real-time (RT) processing and is typically RAN Slice type specific after the completion of the UE access procedure.
- PHY denotes the physical layer processing per RAN Slice and is working either OnPrem or on FarEdge and performs the L1-Hi (layer 1 higher layer function) for RAN slice specific functions. The OnPrem PHY additionally performs L1-Hi for synchronization, cell broadcast and the access procedure.

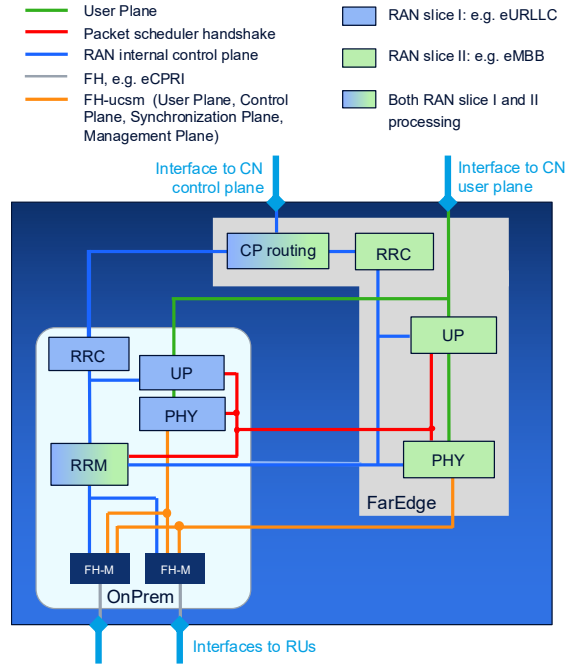

**FIGURE 4.** 6G RAN - Example deployment for two RAN slices.

- Fronthaul-Multiplexer (FH-M) is needed per Radio Unit (RU) and resides OnPrem cloud only. The FH-M interfaces to one single RU and performs FH-user, control, synchronization and management traffic multiplexing and Routing for DL and UL.

### D. OPTIMIZING THE PROTOCOL STACK

The overall target for the 6G protocol stack is to improve its efficiency, i.e., lowering the operating costs while still maintaining the flexibility for diverse QoS treatments for the expected use-cases. That is achieved by simplification and one important aspect of this simplification is the interplay of the different functionalities in the user plane protocol stack, which in 5G is largely kept separated on the different layers MAC, RLC, PDCP.

One example for such functionality is pre-processing in relation to queue management. Pre-processing of incoming packets should be enabled to finalize processing-heavy tasks as soon as possible and independent of lower layer transmission opportunities and timings (e.g., encryption, integrity protection, header building, as mainly done on PDCP layer in 5G). Processed packets can thereafter be enqueued in the transmit buffer. Upon a later transmission opportunity, a transport block can then be built based on the pre-processed packet data as fast as possible.

To ensure low latency while achieving full utilization of the radio resources, the transmit buffer must be managed, i.e., balanced between underutilization of the (varying availability of) radio resources vs. overload of the buffer (entailing high packet latency). With active queue management (AQM) in the RANF, congestion indications are sent to the traffic source (transport layer TCP/QUIC) to adjust its transmission rate, thus impacting the managed buffer state. AQM may artificially drop a buffered packet or build on IP explicit congestion notification (ECN) fields for this purpose. While in 5G, transmit buffers may be split between CU (with PDCP) and DU (with RLC) or in DC, for 6G only a single buffer close to the transmitter should be supported. This would be on RLC layer, while PDCP operates buffer-free, i.e., (pre-) processes packets only. At the same time, the congestion indication should not create additional delays. However, as explained above for pre-processing, it is beneficial to apply PDCP sequence number (SN) assignment and ciphering as early as possible, i.e., upon arrival of IP packets, while AQM ideally drops packets from the front of queue for fastest congestion indication, i.e., already SN-assigned/ciphered enqueued packets. This, however, strictly following the 5G standard, introduces a SN gap and leads to reordering in the PDCP receiver and thus to a delayed congestion indication. Also, in 5G, ECN marking would not be possible on front of queue of encrypted IP packets. Therefore, for 6G we propose a Layer 2 indication (e.g., in RLC header) for AQM to overcome the beforementioned challenges. This way, especially critical communication services are better supported, benefitting also from low latency low loss scalable throughput (L4S) [13]. For legacy AQM dropping, a L2 "drop indication" can be introduced to avoid re-ordering delays in the PDCP receiver due to the missing packet, e.g., Rel-18 "SN gap report" in [14].

Moreover, one way to reduce complexity for 5G NR's multi-layer retransmission functionality is to introduce interaction between protocol layers. For example, by making the downlink HARQ's feedback reliable by conveying it via CRC-check layer 2 signaling instead of via the physical layer, the HARQ feedback could readily be reused at the RLC layer to trigger RLC retransmissions [15].

### E. SERVICE ORCHESTRATION

The evolution of 6G networks demands highly flexible and efficient management of RAN resources to meet diverse service requirements. At the core of this evolution lies an E2E orchestration, an enabler for deploying and operating network slices with stringent latency, reliability, and scalability needs. The E2E orchestrator ensures the deployment of network slices by deriving and managing SLAs across the RAN, transport, and core network domains. Coordination between the RAN and transport domain controllers refines SLAs, particularly for fronthaul and mid-haul operations, ensuring E2E SLA fulfillment.

The RAN domain controller manages cloud and network functions essential for network slice deployment and operation. Among its components:
- Cloud Network Function (CNF) Deployment Services: These automate the deployment of RAN components in diverse cloud environments, streamlining tasks like processing resource allocation, configuration, and scaling. By integrating with monitoring tools, they ensure elasticity, version control, and optimized resource usage, pivotal for cloud-native RAN setups.
- Intelligent Functions: these leverage AI/ML to dynamically optimize RAN operations, including radio resource allocation, power control, and interference management. They predict traffic patterns, enhance performance, and support network slicing for varied use cases, while automating tasks through open interfaces like O-RAN's Non-Real-Time and Near-Real-Time RICs.
- Infrastructure Management Services: They handle infrastructure provisioning, monitoring, and maintenance, ensuring scalability, fault tolerance, and cost optimization.

These services enable the RAN to harness cloud computing benefits while maintaining high performance and reliability.

### F. SUB-NETWORKS

Sub-networks are envisioned as one of the components of the 6G-ANNA vision for the Network of Networks [2]. A sub-network is defined as a specialized network to serve localized environments for heterogenous use cases in consumer, automotive or industrial scenarios. A sub-network is typically composed by a sub-network controller serving several sub-network devices, it may be connected to a parent/umbrella network, but it has some level of autonomy, being able to operate also out of the parent network coverage, see Fig. 5.

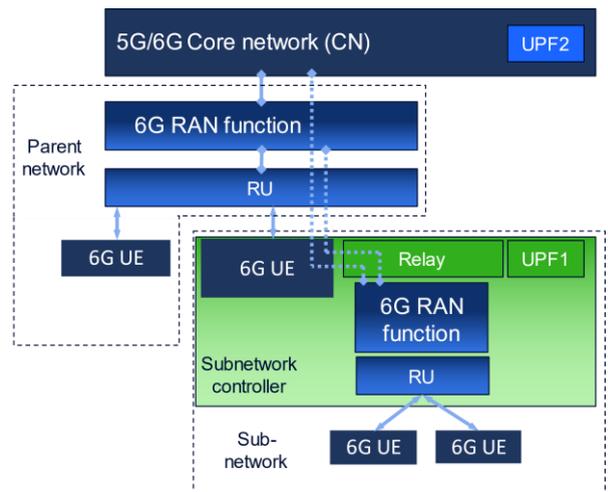

**FIGURE 5.** Sub-network architecture based on Layer 3 relay.

Two types of traffic need to be handled: a) non-local traffic, for example generated by a sub-network device and intended to a user or server outside the sub-network or b) local traffic exchanged among devices within the same sub-network. To handle non-local traffic, the sub-network controller needs to be connected to the parent network and thus it acts as a layer 3 (L3) relay. On the other hand, to handle intra-sub-network communications, the sub-network controller acts as kind of access point, connected to a local core network, and with the possibility of scheduling transmissions from/to/among sub-network devices. Because of that, the sub-network controller is a sort of a network node in-between legacy UEs and gNBs. From the perspective of a gNB, the sub-network controller is a UE with some advanced functionalities and therefore a Uu interface can be assumed between parent network and sub-network controller. Depending on the use case, several possibilities exist on how to define the interface between sub-network devices and sub-network controller: a) for use cases as in/on-body sub-networks, the sub-network controller may be a smartphone and the sub-network devices wearables - such use cases appear as a sort of device-to-device (D2D) scenario, and the interface may be an evolution of 5G NR Sidelink PC5 interface; b) other use cases, for example for in-vehicle, the sub-network controller is more a gNB-like node and the interface within the sub-network may be assumed as Uu. Two RAN-related functionalities that are essential for most of the sub-network use cases are:

- Radio Resource Management, for example with the sub-network controller able to get resources from the parent network and schedule these resources for the intra-sub-network communications.
- Mobility, for example to support handover of a sub-network device from a sub-network to another sub-network or to the parent network.

### G. DISTRIBUTED MIMO

Distributed MIMO refers to a set of techniques for jointly processing multiple spatially distributed RUs as a large, distributed antenna array. These techniques are natively supported by the proposed architecture, where a 6G RANF may serve multiple RUs. Especially if accompanied by RUs densification, and user-centric resource allocation, these techniques can offer uniformly good quality of service to all UEs without the need for expensive, high-powered hardware, enabling more cost-effective and energy-efficient solutions. In addition, these techniques enable seamless handover procedures and simplified simultaneous access from multiple RUs, see also Fig. 6. Even beyond, "cell-free" access can potentially be emulated by an appropriate RANF-RANF interface. Distributed MIMO can be seen as an evolution of older Coordinated Multi-Point (CoMP) concepts, which leverages the recent innovations in massive MIMO, network disaggregation, and virtualization.

Despite the benefits, distributed MIMO comes with less obvious architectural trade-offs compared to conventional cellular connectivity from a single RU. Key challenges include identifying the most efficient LLS for fundamental procedures such as channel estimation, beamforming, and resource allocation, i.e., PHY sub-functions that may be executed at the RU. For instance, recent research indicates that leveraging the local processing capabilities of the RUs for computing at least a portion of the beamformers can significantly reduce fronthaul and computational requirements compared to solutions where beamformers are entirely computed by the gNB (i.e., RANF in 6G). Another critical aspect is determining the appropriate temporal granularity for configuring the various transmission parameters. In fact, conventional resource allocation techniques that operate on a small-scale fading basis are likely impractical in large, distributed MIMO systems due to excessive signaling overhead and computational complexity. Therefore, alternative solutions that optimize network parameters based on slowly varying macroscopic statistical channel properties become highly attractive, especially when they result in only minor performance degradation.

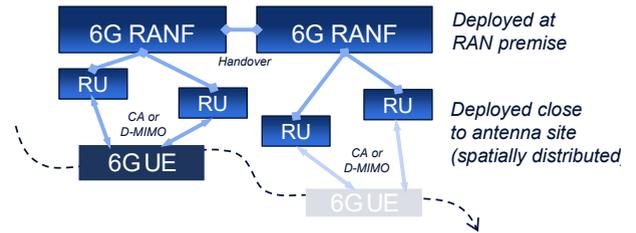

**FIGURE 6.** Deployment of CA or D-MIMO with distributed RUs.

The answers to these questions are highly dependent on specific deployment characteristics and operational regimes, for instance in terms of number of RU antennas, RU density, mobility patterns, propagation conditions, fronthaul capacity, and processing power. Therefore, for the network to be effective, distributed MIMO must be combined with other solutions that enable flexible function placement according to the specific needs. This is well aligned with the identified 6G RAN design goals, since the required implementation and deployment flexibility is motivated by significant functional and performance benefits.

### H. CARRIER AGGREGATION DEPLOYMENTS

CA has the advantage over DC in being able to optimize resource allocations with the full knowledge of a UE's control and user plane state within the same RANF and the possibility to restrict the UL to only one RANF. DC, on the other hand, requires an UL connection towards both the master and the secondary gNB. One original design goal of DC in 5G was to aggregate resources from spatially separated nodes but came with the complexities and delays of the required inter-node interface between them. With the RAN architecture proposal for 6G, the RANF deployed at RAN premise can connect with multiple geographically distributed RUs and the same advantage can be realized with CA, see also Fig. 6. With a standardized LLS, this is also possible for RUs from different vendors. This is why we propose to solely build on CA for spectrum aggregation in 6G. The central processing in the controlling RANF optimizes mobility, aggregation of spectrum and UL power control among the connected RUs. Between the RANFs, the envisioned 6G architecture supports handover only, i.e., in contrast to the 5G architecture that supported also DC between multiple gNBs. In comparison to the proposed CA-based architecture this would however only add unnecessary complexity.

*I. ZERO-TRUST ACCESS CONTROL*

Towards a trustworthy 6G RAN, it is widely recommended to apply the zero-trust architecture (ZTA) where every UE requesting for radio access must be carefully assessed according to the trust evaluation [16]. The core of such a trustworthy access control is the Level of Trust Assessment Function (LoTAF), which serves as the trust evaluation engine, providing a neutral and bidirectional service that caters to both trustors and trustees [17]. According to [17], LoTAF is supposed to be designed as a neutral service that could be deployed, e.g., in a decentralized marketplace where on-demand resources and services may be provided. This suggests a distributed deployment across the CN and the non RT RICs.

## V. CONCLUSIONS AND FURTHER WORK

In this paper we presented a RAN architecture for the upcoming 6G radio access network. We addressed the limitations of the current 5G in handling future diverse traffic demands in an energy and operationally efficient manner. For 6G, our goal is to provide a simplified, trustworthy and flexible solution, based on automation and Cloud RAN, and to improve this efficiency. To achieve this, we proposed to streamline the protocol stack and RAN architecture by focusing on important open interfaces and to remove redundant options. Examples focused on one carrier aggregation scheme, sub-networks concepts, orchestration approaches, enhanced security through a zero-trust framework, distributed massive MIMO and network disaggregation. We supported our architecture findings by giving several deployment examples for RAN slices, sub-network integration and carrier aggregation.

In the future research work, we will among others delve deeper into RAN architecture options, e.g., their potential evolution into a service-based architecture. This could be a first step towards a Global Service-based Architecture (GSBA), foreseen by the SNS JU 6G Architecture Working Group [3] to encompass all mobile network domains in the future. Moreover, while current RAN control plane technology lacks slicing capabilities, as future advancements we will analyze also slicing capabilities in the control plane for user prioritization.

## REFERENCES

[1] J. Almodovar and C. Vinagre, eds., "6G Use Cases and Requirements", Hexa-X-II D1.2, Deliverable D1.2, Dec. 2023. [Online]. Available: https://hexa-x-ii.eu/wp-content/uploads/2024/01/Hexa-X-II_D1.2.pdf.

[2] M. Hoffmann et al., "A Secure and Resilient 6G Architecture Vision of the German Flagship Project 6G-ANNA," IEEE Access, vol. 11, pp. 102643-102660, 2023. DOI: 10.1109/ACCESS.2023.3313505. [Online]. Available: http://dx.doi.org/10.1109/ACCESS.2023.3313505.

[3] Smart Networks and Services Joint Undertaking (SNS JU) 6G Architecture Working Group, "Towards 6G Architecture: Key Concepts, Challenges, and Building blocks", pre released Mar. 2025. [Online]. Available: https://ezywureyi7i.exactdn.com/wp-content/uploads/2025/03/archwg-whitepaper-v1.3-for-public-consultation.pdf, Accessed on: Apr. 29, 2025.

[4] G. Kunzmann et al., "Technology innovations for 6G system architecture," Nokia White Paper. Nokia Bell Labs: Apr. 2022. [Online]. Available: https://www.nokia.com/asset/212403.

[5] M. A. Habibi et al., "Toward an Open, Intelligent, and End-to-End Architectural Framework for Network Slicing in 6G Communication Systems," IEEE Open Journal of the Communications Society, vol. 4, pp. 1615-1658, 2023. DOI: 10.1109/OJCOMS.2023.3294445.

[6] 3GPP 6G Workshop, 6GWS-250243, Summary of the 3GPP 6G workshop, Incheon, Korea, Mar. 10, 2025. [Online]. Available: https://www.3gpp.org/ftp/workshop/2025-03-10_3GPP_6G_WS/Docs/6GWS-250243.zip. Accessed on: May 5, 2025.

[7] B. Masood Khorsandi et al, "The 6G Architecture Landscape - European perspective," 5GPPP Architecture WG whitepaper, Dec. 2022. [Online]. Available: https://doi.org/10.5281/zenodo.7313232. Accessed on: May 5, 2025.

[8] Nokia, 3GPP workshop on 6G, 6GWS-250004, "6G Radio and RAN", Incheon, Korea, Mar. 10-11, 2025. [Online]. Available: https://www.3gpp.org/ftp/workshop/2025-03-10_3GPP_6G_WS/Docs/6GWS-250004.zip.

[9] Ericsson, 3GPP workshop on 6G, 6GWS-250084, "Overall vision & priorities for RAN in 6G", Incheon, Korea, Mar. 10-11, 2025. [Online]. Available: https://www.3gpp.org/ftp/workshop/2025-03-10_3GPP_6G_WS/Docs/6GWS-250084.zip.

[10] Ericsson, Co-creating a cyber-physical world, white paper, Jul. 2024. [Online]. Available: https://www.ericsson.com/en/reports-and-papers/white-papers/co-creating-a-cyber-physical-world.

[11] O-RAN Working Group 1, O-RAN architecture description 12.00, O-RAN.WG1.OAD-R003-v12.00 Technical Specification, Jun. 2024.

[12] International Telecommunication Union, Recommendation ITU-R M.2160-0 M Series: Mobile, radiodetermination, amateur and related satellite services, Framework and overall objectives of the future development of IMT for 2030 and beyond, Nov. 2023.

[13] Internet Engineering Task Force (IETF), RFC 9330, Low Latency, Low Loss, and Scalable Throughput (L4S) Internet Service: Architecture, Dec. 2023.

[14] 3GPP, TS 38.323 V. 18.3, Packet data convergence protocol specification, Sep. 2024.

[15] M. Phan, M. Abad, T. Dudda, E. Eriksson, J. Skördeman, "User Plane Radio Protocol Concepts for 6G", EuCNC & 6G Summit – WOS, Poszan, Poland, Jun. 2025.

[16] X. Chen, W. Feng, N. Ge, and Y. Zhang., "Zero trust architecture for 6G security," IEEE Network, vol. 38, no. 4, pp. 224-232, Jul. 2024, DOI: 10.1109/MNET.2023.3326356.

[17] B. Masood Khorsandi et al., "Enabling Hexa-X 6G vision: An end-to-end architecture," 2024 Joint European Conference on Networks and Communications & 6G Summit (EuCNC/6G Summit), Antwerp, Belgium, 2024, pp. 676-681, DOI: 10.1109/EuCNC/6GSummit60053.2024.10597113.

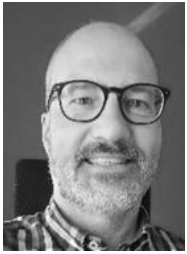
**PETER SCHEFCZIK** is a member of technical staff at Bell Labs, Germany. Dr. Schefczik concentrates his current research interests on laying the foundations for cloud based radio access networks for the future of the mobile network infrastructure. He holds several international patents in the data transmission and mobile networks area and has also authored or co-authored numerous research papers in these fields.

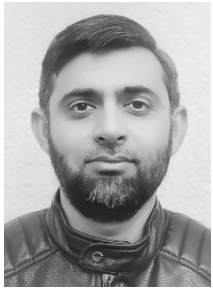
**UMAR TOSEEF** received his Ph.D. (Summa Cum Laude) in Electrical and Communication Engineering from the University of Bremen, Germany, in 2013. He is currently a Senior Wireless Research Engineer at Nokia Bell Labs in Stuttgart, Germany, where he focuses on resource management and virtualization in Cloud RAN. His research interests include software-defined networking (SDN), network function virtualization (NFV), heterogeneous wireless networks, and 6G Network Architecture. Dr. Toseef has extensive experience in EU research projects and holds multiple patents in the field of mobile and wireless communications. He has authored numerous journal and conference papers and received multiple awards for his academic contributions, including two "Best Ph.D. Thesis" honors.

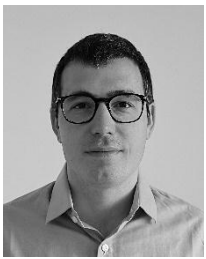
**PAOLO BARACCA** is a Senior Staff Research Specialist at Nokia in Munich, Germany, since 2022. He received the B.Sc. and the M.Sc. degrees in Telecommunications Engineering in 2007 and 2009, respectively, and the Ph.D. degree in Information Engineering in 2013, all from the University of Padova, Italy. After the Ph.D. he joined as a Research Engineer Bell Labs (which was part of Alcatel-Lucent until 2015, and of Nokia since then) in Stuttgart, Germany. He has been an active contributor in several European and German funded research projects on 5G and 6G, including METIS, mmMAGIC, ONE5G and 6G-ANNA. He has also been an active contributor to 3GPP standards development. His research interests include signal processing, multi-antenna techniques and scheduling for wireless communications. He has co-authored more than 50 journal or conference papers and holds more than 150 granted or pending patents.

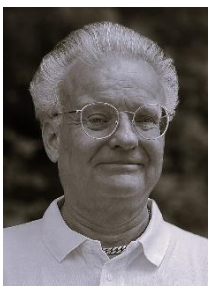
**RALF KLOTSCHE** is senior research engineer since 1988 in Germany. In the past seventeen years at Nokia Bell Labs he focused on distributed, and cloud based real-time solutions for initially multimedia applications and later radio network functions. He holds numerous patents and authored or coauthored several research papers.

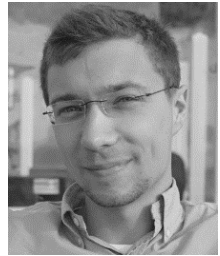
**TORSTEN DUDDA** received a Diploma degree in electrical engineering and information technology from Aachen University, Germany. In 2012, he joined Ericsson in the R&D center Eurolab, close to Aachen. In his current role as Master Researcher, Torsten coordinates teams in internal and external research projects, e.g. in the German funded 6G lighthouse project 6G-ANNA. His current research interests include evolving radio network architectures and protocols toward 6G.

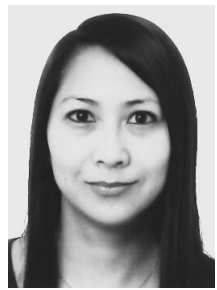
**MAI-ANH PHAN** received her M.Sc. degree in Computer Engineering from RWTH Aachen University in 2007. Afterwards, she immediately joined the Ericsson Eurolab, Germany, as a research engineer. Since then, she was involved in research studies as well as standardization. She worked on various topics ranging from Multimedia Broadcast and Multicast Services in 3G and 4G to shared spectrum channel access in MulteFire and 5G. In her current role as a Senior Researcher, she focuses on the enhancement of radio protocols for 6G.

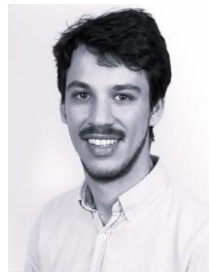
**LORENZO MIRETTI** received the B.Sc. and M.Sc. degrees in Telecommunication Engineering from Politecnico di Torino, Italy, in 2015 and 2018, respectively, and the Ph.D. degree in wireless communications from EURECOM and Sorbonne Université, France, in 2021. He is currently a Senior Researcher with Ericsson Research, Germany. Previously, he held appointments as a post-doctoral researcher with the Technical University of Berlin, Germany, and as a Research Associate with the Fraunhofer Heinrich Hertz Institute, Berlin, Germany. His research interests include communication theory and signal processing for wireless networks, with particular focus on next-generation multi-antenna systems.

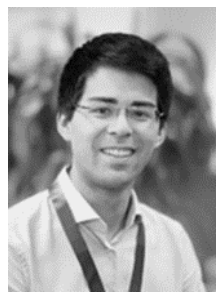
**DAVID GINTHÖR** received his M.Sc. degree in Electrical Engineering and Information Technology from TU Munich in 2018. He joined Robert Bosch GmbH Corporate Research in Stuttgart, Germany, in 2018 as a research engineer. His work focuses on wired and wireless communication in the automotive and industrial domain including Industrial Ethernet, Time Sensitive Networking, and 5G/6G mobile communication.

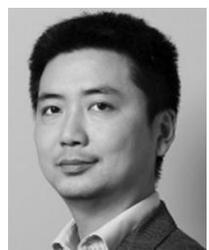
**BIN HAN** received the B.E. degree from Shanghai Jiao Tong University in 2009, the M.Sc. degree from the Darmstadt University of Technology in 2012, and the Ph.D. (Dr.-Ing.) degree from the Karlsruhe Institute of Technology in 2016. He joined the University of Kaiserslautern (RPTU) in July 2016. In November 2023, he was granted the teaching license (Venia Legendi) by RPTU and therewith became a Privatdozent. He is the author of two books, six book chapters, and over 80 research papers. He has participated in multiple EU FP7, Horizon 2020, and Horizon Europe research projects. He is an Editorial Board Member for Network and the Guest Editor for Electronics. He has served in organizing committee and/or TPC for IEEE GLOBECOM, IEEE ICC, EuCNC, European Wireless, and ITC. He is actively involved in the IEEE Standards Association Working Groups P2303, P3106, and P3454, respectively.